\documentclass[aps,twocolumn,amsmath,amssymb]{revtex4}
\usepackage{graphicx}% Include figure files
\usepackage{stmaryrd}
\usepackage{bm}% bold math

\begin{document}

\title{Thorium-doping induced superconductivity up to 56 K in Gd$_{1-x}$Th$_{x}$FeAsO}

\author{Cao Wang, Linjun Li, Shun Chi, Zengwei Zhu, Zhi Ren, Yuke Li, Yuetao Wang, Xiao Lin, Yongkang Luo, Shuai Jiang, Xiangfan Xu, Guanghan Cao\footnote[1]{Electronic address: ghcao@zju.edu.cn} and Zhu'an Xu\footnote[2]{Electronic address: zhuan@zju.edu.cn}}
\affiliation{Department of Physics, Zhejiang University, Hangzhou 310027, People's Republic of China}

\maketitle

\textbf{Following the discovery of superconductivity in an iron-based arsenide LaO$_{1-x}$F$_{x}$FeAs with a
superconducting transition temperature ($T_c$) of 26 K\cite{Kamihara08}, $T_c$ was pushed up surprisingly to
above 40 K by either applying pressure\cite{Takahashi} or replacing La with Sm\cite{Chen-Sm}, Ce\cite{Chen-Ce},
Nd\cite{Ren-Nd} and Pr\cite{Ren-Pr}. The maximum $T_c$ has climbed to 55 K, observed in
SmO$_{1-x}$F$_{x}$FeAs\cite{Ren-Sm,Chen-Sm2} and SmFeAsO$_{1-x}$\cite{Ren-Sm2}. The value of $T_c$ was found to
increase with decreasing lattice parameters in LnFeAsO$_{1-x}$F$_{x}$ (Ln stands for the lanthanide elements) at
an apparently optimal doping level. However, the F$^{-}$ doping in GdFeAsO is particularly
difficult\cite{Chen-elements,Wen-Gd} due to the lattice mismatch between the Gd$_2$O$_2$ layers and Fe$_2$As$_2$
layers. Here we report observation of superconductivity with $T_c$ as high as 56 K by the Th$^{4+}$ substitution
for Gd$^{3+}$ in GdFeAsO. The incorporation of relatively large Th$^{4+}$ ions relaxes the lattice mismatch,
hence induces the high temperature superconductivity.}

LnFeAsO family\cite{Quebe} crystalline in tetragonal ZrCuSiAs-type\cite{Johnson&Jeitschko} structure with space
group \emph{P4/nmm}. From the crystal chemistry point of view, the crystal structure can be described as an
alternate stacking of Ln$_2$O$_2$ fluorite-type block layers and Fe$_2$As$_2$ antifluorite-type block layers
along $c$-axis. The two block layers are connected by CsCl-type layers (Figure 1). Therefore, the chemical
stability of LnFeAsO depends, to some extent, on the lattice match between the two block layers. A rough
estimate based on the effective ionic radii\cite{Shannon} indicates that the Ln$_2$O$_2$ planar lattice is
substantially smaller than the Fe$_2$As$_2$ lattice. The lattice mismatch becomes more serious for the LnFeAsO
member with a smaller Ln$^{3+}$ ion. That may explain why TbFeAsO, DyFeAsO and other heavy-lanthanide-containing
LnFeAsO members were hard to synthesize previously\cite{Quebe}.

\begin{figure}
\includegraphics[width=9cm]{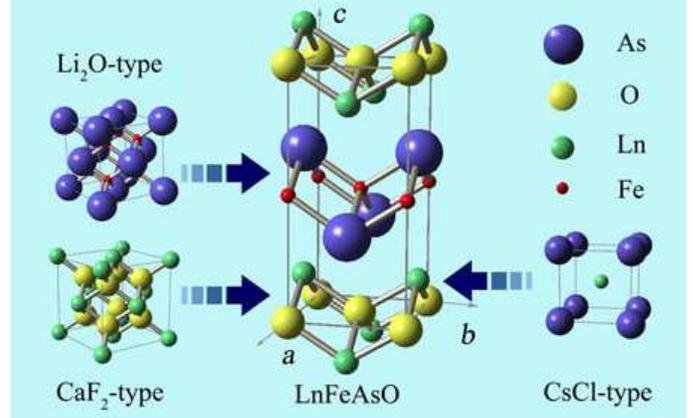}
\caption {\textbf{Crystal chemistry understanding of the structure of LnFeAsO (Ln=lanthanides).} The stacking of
fluorite (CaF$_2$) layers, CsCl-type layers and antifluorite (Li$_2$O) layers along $c$-axis forms the LnFeAsO
structure. The lattice constant along the stacking direction can be expressed by the formula $c \simeq
\frac{1}{2}a_{CaF_{2}}+\frac{1}{2}a_{CsCl}+\frac{1}{2}a_{Li_{2}O}$, which basically satisfies the experimental
result. Note that the lattice mismatch between the Ln$_2$O$_2$ layers and the Fe$_2$As$_2$ layers affects the
chemical stability of LnFeAsO.}
\end{figure}

As the family member with the relatively small Ln$^{3+}$ ion, GdFeAsO is a promising parent compound to have a
higher $T_c$ through carrier doping. Similar to cuprate superconductors in which superconductivity emerges when
charge carriers are induced into CuO$_2$ planes by chemical doping at "charge reservoir layers",\cite{Cava}
superconductivity in LnFeAsO$_{1-x}$F$_{x}$ is realized by partial substitution of O$^{2-}$ with F$^{-}$. The
F$^{-}$-for-O$^{2-}$ substitution introduces extra positive charges in the insulating Ln$_2$O$_2$ layers and
negative charges (electron doping) in the Fe$_2$As$_2$ layers. Earlier preliminary experiment showed sign of
superconductivity below 10 K in GdFeAsO$_{1-x}$F$_{x}$.\cite{Chen-elements} Later, $T_c$ was pushed up to 36 K
in a polycrystalline sample with a nominal composition of GdO$_{0.83}$F$_{0.17}$FeAs.\cite{Wen-Gd} Very recently
the $T_c$ value was increased to 53.5 K in oxygen-deficient GdFeAsO$_{1-x}$ by using high-pressure
synthesis.\cite{Ren-Gd} It is of great interest whether the $T_c$ can be elevated further in electron-doped
GdFeAsO systems.

Up to now, electron doping in the iron-based oxyarsenides was realized through the chemical substitution only at
oxygen site in Ln$_2$O$_2$ layers. Because the ionic radius of F$^-$ (1.31 \AA, CN=4) is distinctly smaller than
that of O$^{2-}$ (1.38 \AA, CN=4)\cite{Shannon}, F$^-$ substitution for O$^{2-}$ in GdFeAsO leads to more severe
lattice mismatch as mentioned above. In other word, doping F$^-$ (or oxygen vacancy) in GdFeAsO is particularly
difficult, which is probably the main obstacle to elevate $T_c$. Substitution of Ln$^{3+}$ by relatively large
tetravalence ions is an alternative route to introduce electrons. A successful example was the electron doping
in Ln$_{2-x}$Ce$_{x}$CuO$_4$ (Ln=Pr, Nd or Sm), which has led to the discovery of $n$-type cuprate
superconductors\cite{Tokura89}. Th$^{4+}$ is a very stable tetravalence ions and is as large as
Gd$^{3+}$\cite{Shannon}, therefore, we pursued the Th$^{4+}$ substitution for Gd$^{3+}$ in the GdFeAsO system.

Figure 2a shows the X-ray diffraction (XRD) patterns of the Gd$_{1-x}$Th$_{x}$FeAsO samples. The XRD peaks of
the undoped compound can be well indexed based on the tetragonal ZrCuSiAs-type structure, indicating single
phase of GdFeAsO. As for the Th-doped samples, small amount of unreacted ThO$_2$ can be seen. The refined
lattice parameters are $a$ = 3.9154(2) {\AA} and $c$ = 8.4472(4) {\AA} for the parent compound, basically
consistent with the previously reported values\cite{Quebe}. For the Th-doped sample of $x$=0.2, the fitted
lattice parameters are $a$ = 3.9161(2) {\AA} and $c$ = 8.4386(3) {\AA}. The other Th-doped sample of $x$=0.25
has a similar unit cell with $a$=3.9166(3) {\AA} and $c$ = 8.4382(5) {\AA}. Therefore, it is evident that the
Th-doping tends to expand the lattice within $ab$-planes but to shorten the lattice along the $c$-axis. The
substantial change in cell constants indicates that thorium is incorporated in the lattice. Meanwhile, the
$ab$-plane expansion reflects the effective electron doping onto FeAs layers.

More direct evidence of Th incorporation into the lattice comes from the chemical composition measurement by
energy-dispersive X-ray (EDX) microanalysis. Figure 2b shows that the microcrystal in the SEM image contains
remarkably Th in addition to Gd, O, Fe and As. Quantitative analysis gives the Gd:Th:Fe:As ratios as
0.83:0.19:0.96:1.00 for the $x$=0.2 sample (Here we omit the oxygen content because the amount of oxygen cannot
be measured so precisely by EDX technique). This result demonstrates that most of the Th was successfully doped
for the sample of $x$=0.2. Obviously, Th-doping level is higher than that of F-doping in
GdFeAsO\cite{Chen-elements,Wen-Gd}.

\begin{figure}
\includegraphics[width=7cm]{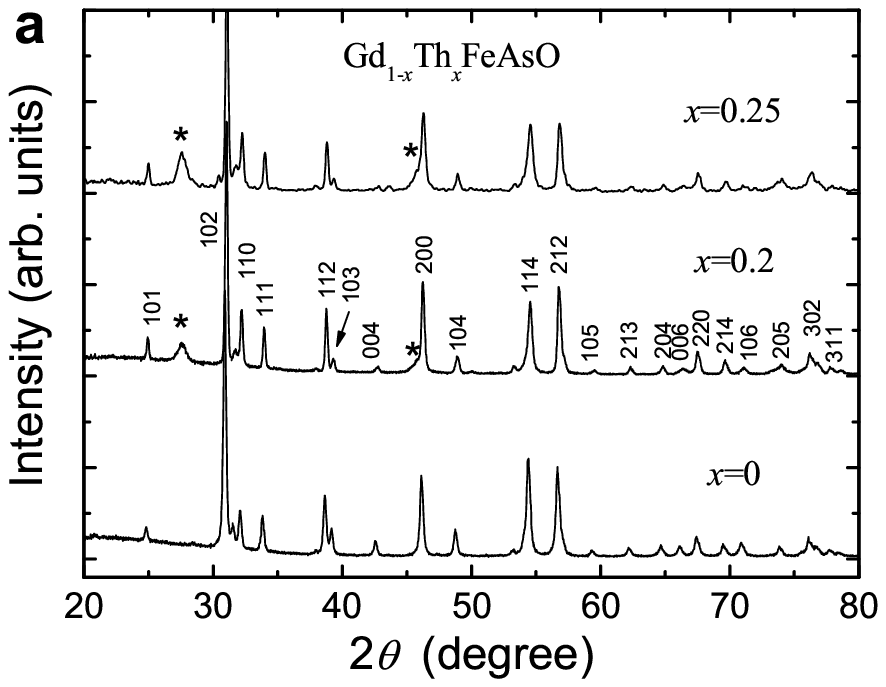}
\includegraphics[width=7cm]{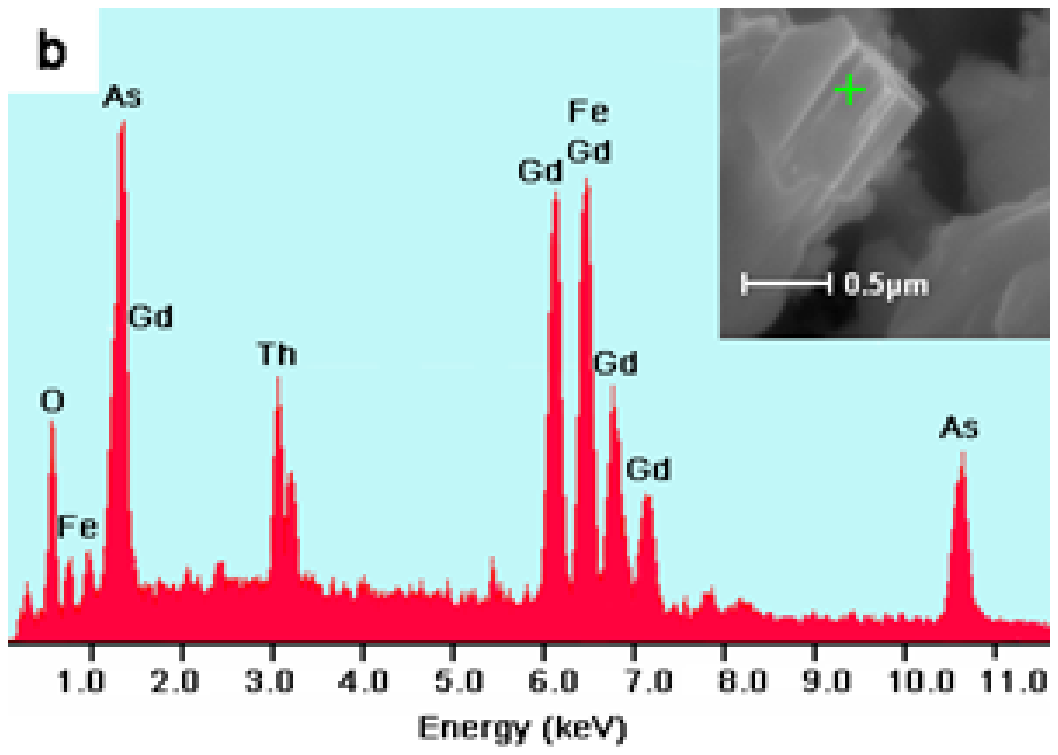}
\caption {\textbf{Phase and compositional analysis of Gd$_{1-x}$Th$_{x}$FeAsO samples.} \textbf{a}, Powder XRD
of Gd$_{1-x}$Th$_{x}$FeAsO polycrystalline samples. The asterisked peaks come from unreacted ThO$_{2}$.
\textbf{b}, a representative EDX spectrum of a crystalline grain of Gd$_{0.8}$Th$_{0.2}$FeAsO sample, indicating
the incorporation of thorium into the lattice. The inset shows the SEM image of the identical sample. The marked
spot is the position where the EDX spectrum was collected.}
\end{figure}

Figure 3a shows the $\rho(T)$ curve for Gd$_{1-x}$Th$_{x}$FeAsO samples. For the undoped GdFeAsO, the $\rho(T)$
curve exhibits an obvious anomaly at 128 K, characterized by a resistivity drop with decreasing temperature. In
LaFeAsO a similar resistivity anomaly was found at 150 K,\cite{Kamihara08,wnl} which has been recently suggested
to be associated with a structural phase transition and/or an antiferromagnetic spin-density-wave
transition\cite{wnl,Cruz,Mcguire,Nomura}. We speculate that the present resistivity anomaly in GdFeAsO has a
similar physical origin with that in LaFeAsO. For $x$=0.2 and $x$=0.25, such resistivity anomaly disappears,
instead, resistivity drops abruptly to zero below 55 K, indicating a superconducting transition. In addition,
the linear temperature-dependence of normal-state resistivity near $T_c$ suggests possible non-Fermi liquid
behavior in the present system, similar to that in SmFeAsO$_{1-x}$F$_{x}$\cite{Chen-Sm2}.

\begin{figure}
\includegraphics[width=7cm]{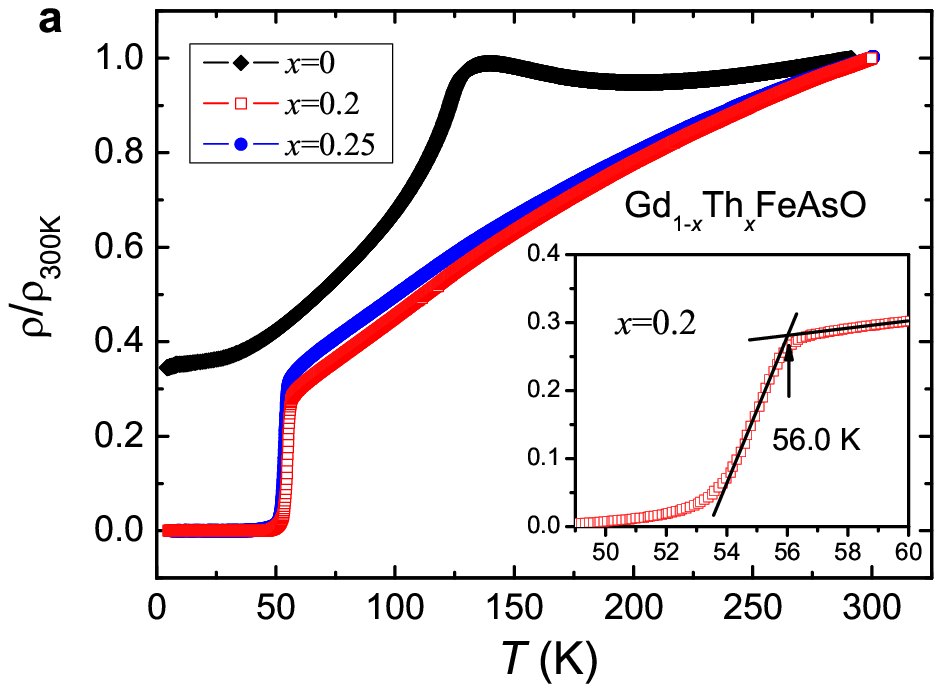}
\includegraphics[width=7cm]{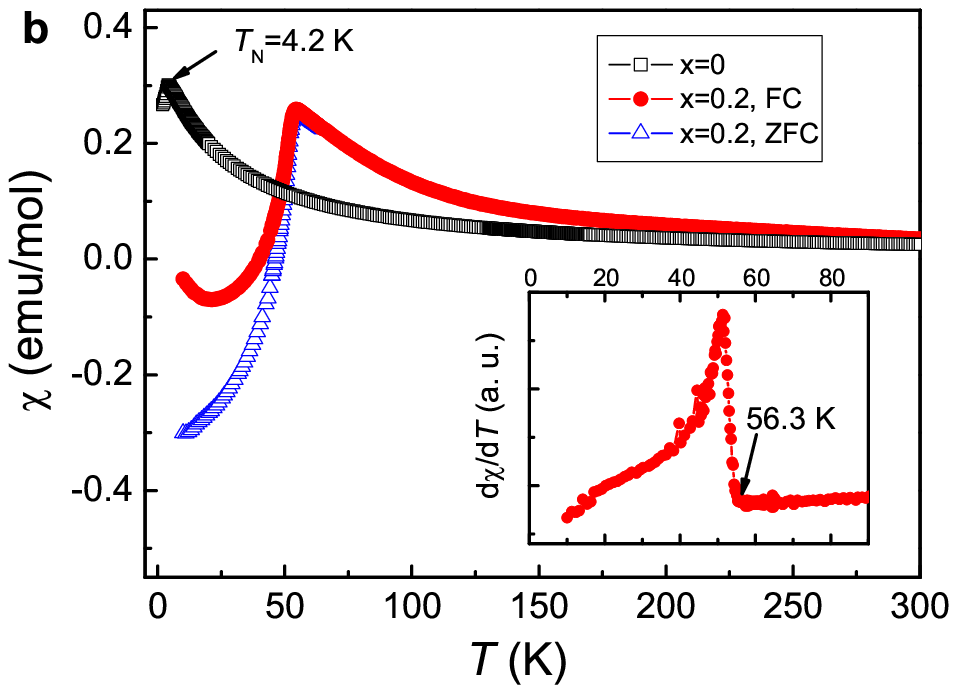}
\caption {\textbf{Electrical resistivity ($\rho$) and magnetic susceptibility ($\chi$) of
Gd$_{1-x}$Th$_{x}$FeAsO.} \textbf{a}, $\rho$ versus $T$. The data are normalized to $\rho_{300K}$ as the
resistivity measured on polycrystalline samples is often higher than the intrinsic value due to the grain
boundary and surface effect. The inset is an expanded plot to show the superconducting transition at 56 K for
the sample of $x$=0.2. \textbf{ b}, $\chi$ versus $T$. The lower inset shows the differential $\chi_{FC}$ curve
of Gd$_{0.8}$Th$_{0.2}$FeAsO powder sample, which locates the onset $T_{c}$ at 56.3 K. ZFC, zero-field cooling;
FC, field cooling.}
\end{figure}

The magnetic measurement on Gd$_{1-x}$Th$_{x}$FeAsO samples was shown in Figure 3b. In the parent compound, the
high-temperature magnetic susceptibility ($\chi$) follows the Curie-Weiss law. The fitted effective magnetic
moments was 7.96 $\mu_B$ per formula unit, in good agreement with the magnetic moment of free Gd$^{3+}$ ion.
Below 4.2 K, $\chi$ drops sharply, indicating an antiferromagnetic ordering of Gd$^{3+}$ magnetic moments. Note
that similar behavior was observed in SmFeAsO where the Neel temperature was 4.6 K.\cite{lsy}

For the Th-doped samples, e.g., $x$=0.2, the normal-state susceptibility roughly obeys Curie-Weiss law. Below 55
K, $\chi$ drops steeply to negative values, confirming the superconductivity observed above. After subtracting
the paramagnetic susceptibility of Gd$^{3+}$ ions, the volume fraction of magnetic shielding at 10 K was
estimated to be over 50\%, indicating bulk superconductivity. The differential $\chi_{FC}$ curve in the inset of
Figure 3b shows that the $T_{c}$(onset) is over 56 K, consistent with the resistance measurement. Therefore,
Th-doping in GdFeAsO elevates the $T_{c}$ by 20 K ($T_{c}$=36 K in GdO$_{0.83}$F$_{0.17}$FeAs\cite{Wen-Gd}).
Moreover, the $T_c$ of 56 K is among the highest ever discovered in iron-based oxypnictides.

Our observation of superconductivity in Gd$_{1-x}$Th$_{x}$FeAsO indicates that the Ln-site substitution in
LnFeOAs is feasible to realize electron doping, hence the high temperature superconductivity. Moreover, the
electron-doping is more easily induced by the Th$^{4+}$-substitution compared with the F$^{-}$-substitution in
GdFeOAs. This manifests that the lattice match between Ln$_2$O$_2$ fluorite-type layers and Fe$_2$As$_2$
antifluorite-type layers is important not only for the chemical stability of the parent compounds, but also for
the electron-doping level in LnFeAsO. It is thus expected that the thorium-doping strategy can be applied to
other iron-based oxypnictides.

\textbf{METHODS}

--------------------------------------------------------

\textbf{Sample preparation}

Polycrystalline samples of Gd$_{1-x}$Th$_{x}$FeAsO were synthesized by solid state reaction in an evacuated
quartz tube. All the starting materials (Gd, Gd$_{2}$O$_{3}$, ThO$_{2}$, Fe and As) are with high purity ($\geq$
99.95\%). First, GdAs was presynthesized by reacting Gd tapes with As pieces in vacuum at 773 K for 10 hours and
then 1173 K for 12 hours. Similarly, FeAs was prepared by reacting Fe powders with As shots at 773 K for 6 hours
and then 1030 K for 12 hours. Then, powders of GdAs, Gd$_{2}$O$_{3}$, ThO$_{2}$, Fe and FeAs were weighed
according to the stoichiometric ratio of Gd$_{1-x}$Th$_{x}$FeAsO. The weighed powders were mixed thoroughly by
grinding, and pressed into pellets under a pressure of 4000 kg/cm$^{2}$ in an argon-filled glove box. The
pressed pellets were wrapped with Ta foils, and sealed in an evacuated quartz ampoule. The sealed ampoule was
slowly heated to 1423 K, holding for 48 hours. Finally the samples were rapidly cooled to room temperature.

\textbf{Structural characterizations}

Powder X-ray diffraction was performed at room temperature using a D/Max-rA diffractometer with Cu K$_{\alpha}$
radiation and a graphite monochromator. The XRD diffractometer system was calibrated using standard Si powders.
Lattice parameters were refined by a least-squares fit using at least 20 XRD peaks. Energy-dispersive X-ray
(EDX) spectra were obtained by using the Phoenix EDAX equipment attached to a field-emission scanning electron
microscope (SIRION FEI). The ground sample powders were placed directly on the copper sample holder for making
the SEM specimen.

\textbf{Electrical and magnetic measurements}

The electrical resistivity was measured with a standard four-terminal method. Samples were cut into a thin bar
with typical size of 4mm$\times$2mm$\times$0.5mm. Gold wires were attached onto the samples' newly-abraded
surface with silver paint. The size of the contact pads leads to a total uncertainty in the absolute values of
resistivity of ¡À10 \%. The electrical resistance was measured using a steady current of 5 mA, after checking
the linear $I-V$ characteristic.

Temperature dependence of magnetization was measured on a Quantum Design Magnetic Property Measurement System
(MPMS-5). For the measurement of the undoped compound, the applied field was 1000 Oe. For the measurement of the
Th-doped superconducting samples, both the zero-field-cooling and field-cooling protocols were employed under
the field of 10 Oe.

%References

\begin{acknowledgments}
The authors thank Fuchun Zhang, Ying Liu and Xianhui Chen for helpful discussions. This work is supported by the
National Basic Research Program of China (No.2006CB601003 and 2007CB925001) and the PCSIRT of the Ministry of
Education of China (IRT0754).
\end{acknowledgments}
\end{document}